\documentclass[twocolumn,showpacs,preprintnumbers,amsmath,amssymb]{revtex4}

\usepackage{graphicx}
\usepackage{dcolumn}
\usepackage{bm}

\begin{document}

\preprint{submitted to Phys. Rev. Lett.}

\title{
A Novel Photonic Material for Designing Arbitrarily Shaped Waveguides in Two Dimensions}

\author{Hiroshi Miyazaki}
 \email{hmiyazak@olive.apph.tohoku.ac.jp}
\author{Masashi Hase$^{1}$}
\author{Hideki T. Miyazaki$^{1}$}
\author{Yoichi Kurokawa}
\author{Norio Shinya$^{1}$}
\affiliation{%
Department of Applied Physics, Tohoku University, Aoba, Aramaki-aza,
 Aoba-ku, Sendai, Miyagi, 980-8579, Japan
\\$^{1}$National Institute for Materials Science (NIMS), 1-2-1 Sengen,
 Tsukuba, 305-0047, Japan
}

\date{\today}

\begin{abstract}

We investigate numerically optical properties of novel two-dimensional photonic materials where parallel dielectric rods are randomly placed with the restriction that the distance between rods is larger
 than a certain value.
A large complete photonic gap (PG) is found when rods have sufficient
 density and dielectric contrast. 
Our result shows that neither long-range nor short-range order is
 an essential prerequisite to the formation of PGs. 
A universal principle is proposed for designing arbitrarily
 shaped waveguides, where waveguides are fenced with side walls of
 periodic rods and surrounded by the novel photonic materials. 
We observe highly efficient transmission of light for various
 waveguides.
Due to structural uniformity, the novel photonic materials are best
 suited for filling up the outer region of waveguides
 of arbitrary shape and dimension comparable with the wavelength. 
 
\end{abstract}


\maketitle

Downsizing is an everlasting dream of researchers in engineering field. 
Researchers in the field of optics are hoping to find a way to fabricate
 all-optic integrated circuits by using optical elements comparable
 with the wavelength. 
In fact, realization of miniature-sized optical waveguides should soon
 be possible due to the discovery of photonic crystals (PhCs)
 \cite{JoannopoulosNature97, LinScience98}. 
Because of the periodicity of a dielectric constant, PhCs can be designed
 to have complete photonic gaps (PGs), range of frequencies for which
 light in any direction cannot propagate within the PhCs. 
We can steer light through the waveguides made of PhCs having complete PGs. 
On the other hand, the shapes of waveguides should be structurally
 commensurate with the periodicity of the host PhC. 
This severely obstructs the realization of arbitrarily shaped waveguides
 whose dimension is comparable with the wavelength.

Let us take two-dimensional waveguides of a PhC composed of periodic
 dielectric rods, for example. 
The waveguides are usually formed by removing rods along a line. 
Therefore, they are composed of a set of segments. 
Waveguides of $90^\circ$ or $60^\circ$ bends can be easily obtained from PhCs of
 square or triangular lattices. 
For a bend of an arbitrary angle, however, they become zigzag in shape
 and excess scattering occurs at the junctions of segments. 
Successive scattering significantly reduces the transmittance of
 waveguides composed of many branches and bends. 
Arbitrarily shaped waveguides, therefore, require photonic materials of
 maximum structurally uniformity in addition to the complete PGs.
In this Letter, we propose novel photonic materials in which parallel
 dielectric rods are randomly placed in a certain region provided that
 the distance between the centers of rods is larger than a certain value $D_{min}$: 
\begin{equation}
|{\bf R}_{i}-{\bf R}_{j}| \geq D_{min},
\end{equation}
where ${\bf R}_{i}$ and ${\bf R}_{j}$ are the positions of $i$-th and $j$-th rod center.
We call this new photonic material as uniformly distributed photonic scatterers (UDPS).
It is noted that UDPS have neither long-range nor short-range order. 
Nevertheless, we show numerically that UDPS can have complete PGs if
 rods have sufficient density and dielectric contrast. 
We also propose a new concept to fabricate arbitrarily shaped waveguides, i. e.,
 we fence the waveguides with side walls of periodic rods and fill up the
 outer region with UDPS. 
We observe clear propagation of waveguide modes with large transmittance.

Two examples of UDPS are shown in Figs. 1a and 1b composed of $N=100$ and
 $N=200$ rods of radius $a$, respectively. 
Here, we put $D_{min}=4.0a$. 
As the figures show, the distribution of rods becomes more
 uniform with increase in rod density. 
Transmittance of UDPS is calculated by assuming the incidence of
 the plane electric field of wavelength $\lambda$ from the upper side
 of Fig. 1a or 1b. 
Incident light is scattered multiply by each rod. 
This scattering is treated analytically by solving the Maxwell equation\cite{Yousif88}.
The solution gives the distribution of electric field and energy flow (Poynting vector). 
From the average energy flow at line L in Fig. 1a or 1b, we calculate
 the transmittance $T$ normalized by that without rods. 
$T$ is a function of normalized frequency $\Omega=2\pi a/\lambda$
 (known as size parameter) and becomes very small in the PG region. 
In some cases, $T$ exceeds unity because of the diffraction due
 to the finite size of UDPS.

Figure 1c shows values of $T$ for UDPS of $N=100$, 150, and 200 when
 the electric field ${\bf E}$ is parallel to the rod axis (TM mode). 
In all the figures of this Letter, we fix the dielectric constant of rods at $\varepsilon=12$
 which corresponds to that of Si at 1.55 $\mu m$ used world-wide in
 optical communications. 
We found no PG for UDPS of $N=100$ though there were two split dips
 at $\Omega=0.35$. 
For UDPS of $N=150$, we observed a PG of $\Delta \Omega /\Omega_c=30\%$,
 where $\Omega_c$ and $\Delta \Omega$ are the central frequency and 
width of the PG, respectively. 
Here, we define PG to be the frequency range continuously below $T=0.01$. 
This PG, however, is incomplete in that there appear spiky peaks
 in the gap region. 
At these spiky peaks we found an intrusion of energy flow through
 cracks (non-uniform regions) of UDPS. 
When $N$ becomes 200, PG grows up to $37\%$ with the suppression of spiky peaks. 
It was also verified that this PG is isotropic and therefore complete.
We also calculated $T$ of UDPS for various $\varepsilon$ with common values of
 $D_{min}=4.0a$ and area fraction of rods $f = 0.138$, and found PG for
 $\varepsilon \geq 5$. 
The presence of PGs was also confirmed by results of finite difference time
 domain (FDTD) calculation. 
Note that the rod radius becomes 0.11 $\mu m$
 if we correspond $\Omega_c$ of $N=200$ to $\lambda=1.55 \mu m$.
This size can be prepared relatively easily using recently developed
 microfabrication technique.

The actual fabrication process inevitably involves certain fluctuation in
 rod position $\Delta x$ and radius $\Delta a$. 
It is natural to expect from the construction rule of UDPS that the PG is
 unaffected by $\Delta x$. 
In contrast, the effect of $\Delta a$ should be investigated. 
We have also plotted in Fig. 1c the transmittance of UDPS with
 $\Delta a/a=\pm 20\%$. 
It was confirmed that PG of $\Delta \Omega /\Omega_c=30\%$ can survive 
 such large fluctuation. 
This means that the PG of UDPS is also considerably robust against radius fluctuation.

We have observed in Fig. 1 that an increase in rod density enlarges the PG width. 
We therefore examined a case of much higher density. 
Figure 2a shows one example of UDPS of $D_{min}=2.1a$, which includes no radius fluctuation. 
Two examples of radial distribution functions $g(r)$ are plotted
 in Fig. 2b for $D_{min}=2.1a$, where $r$ is the distance between
 rod centers. 
For uniform distribution in two dimensions, $g(r)$ is proportional
 to $r$ without showing any peak. 
Distributions in Fig. 2b are very similar to the uniform case. 
Nevertheless, we find three distinct PGs at $\Omega_c$=0.54, 0.93
 and 1.34 for the TM mode in Fig. 2c. 
It is remarkable that the UDPS have such wide PGs of higher frequencies. 
If one uses the third PG, rod radius of $0.33\mu m$ is required to
 utilize $\lambda=1.55\mu m$. 
This facilitates the fabrication significantly. 
We also find a PG of TE mode (${\bf E} \perp$ rod axis)
 at $\Omega_c=0.68$.

Before discussing the origin of PGs in UDPS, let us use UDPS for various
 waveguides. 
For this purpose we plot in Fig. 3 the average transmittance $T$ over five configurations of UDPS
 for three cases of sample thickness with common $D_{min}$. 
We can see that PG appears even for very thin UDPS containing
 three or four rods along the direction of light. 
This indicates that UDPS have wide applicability to build up
 waveguides of arbitrary shape and size comparable with the wavelength.

To make best use of this property, we first decide the shape of a waveguide. 
It can be twisty, as shown in Fig. 4, to enable maximum flexibility in designing. 
Waveguides are separated by side walls from the surrounding medium. 
To avoid excess scattering from the side walls, they are chosen to be made
 of periodic rods in a line.
Then, we fill up the outer region with UDPS. 
It is noted that UDPS are the best materials for the surrounding optical medium. 
PhCs are not suitable for this purpose because their periodicity conflicts
 with that of side walls. 
This mismatch causes non-uniformity of rod density, resulting in excess scattering. 
This is also the case not only for quasi-periodic PhCs \cite{Chan98} but
 also for photonic materials having short-range order \cite{Jin01}.

In the waveguides shown below, the density of UDPS is the same as that
 in Fig. 1b ($N=200$) and radius fluctuation of $\Delta a/a = \pm 20\%$  is introduced. 
Thus, they have a common PG of $0.366 \leq \Omega \leq 0.494$ shown by the
 shaded region in each transmittance.
We assume that the TM mode is incident from the upper side. 
Figure 4a shows the distribution of electric field intensity
 in a waveguide of $90^\circ$ bend. 
Rods are shown by open circles. 
The intensity increases from blue to red. 
Energy flow is shown by white arrows. 
Corresponding frequency is indicated by the arrow in the transmittance
 in Fig. 4b which shows relatively large values over a wide range within PG. 
We can clearly observe the propagation mode of large transmittance
 comparable with that of waveguides made of PhCs. 
We also found no appreciable change in transmittance for a smaller value of $\Delta a$.
Therefore, waveguides of UDPS have a wide tolerance for the fabrication process.

UDPS are not limited to waveguides composed of segments. 
They can also be used for twisty waveguides whose curvatures are comparable with $\lambda$. 
Figure 4c shows such an example composed of two quarter circles. 
The distribution of electric field intensity and energy flow
 are shown at the frequency noted by the arrow in the transmittance in Fig. 4d. 
The rod density and incident light are the same as those in Fig. 4a. 
As can be seen in the figure, the electric field flows smoothly
 downward through the sample. 
While the corresponding value of $T$ is not large (0.739),
 it can be increased by optimization. 

Let us discuss the origin of PGs in UDPS.
In a study concerning effects of disorder on PG, it was found that there
 are two kinds of PG, one that is easily smeared out by disorder and the
 one that is very robust against disorder \cite{Lidorikis00}. 
The former PGs are formed by the coherent interference of
 scattered waves from periodic rods like Bragg diffraction in X-rays. 
The latter are formed by the bonding and anti-bonding states of
 Mie resonance states within each rod, similar to the electronic
 bandgaps in semiconductors. 
Since the latter are formed by local interaction, they are not
 significantly affected by the fluctuations in position and radius. 
As a matter of fact, an isolated dielectric rod of $\varepsilon=12$
 has Mie resonance at $\Omega$=0.23, 0.66, 1.06, 1.14 and 1.44
 for the TM mode and $\Omega$=0.66, 1.05 and 1.40 for the TE mode. 
It is likely that the PG of UDPS is a result of interaction of
 these modes. 
This is also evidenced by the appearance of PG in Fig. 3 for very
 thin UDPS which is easily understood from the formation of
 bonding and anti-bonding states by local arrangement of rods. 
An important difference, however, exists between electrons and
 photons in that resonance wavefunctions of photons are not
 localized exponentially. 
Rather, they decay in inverse power and have a long-range nature. 
This long-range nature is responsible for the formation of PGs
 in UDPS that does not require even a short-range order.

One might think that UDPS is deeply related with two-dimensional disordered systems which are used to investigate Anderson localization of light\cite{John90}.
Let us discuss this point. 
In three dimensions, Anderson localization takes place only when the disorder is strong enough to satisfy the Ioffe-Regel criterion. 
In contrast, even a very small amount of disorder is sufficient in one and two dimensions to invoke Anderson localization.
We have evaluated the localization length $\ell_{loc}$ in various UDPS samples of common
 $D_{min}=4.0a$ and different sample thickness $\ell$. 
We assume the form $T=T_{0} \exp(-\ell/\ell_{loc})$, and
 found that within the PG region
 $\ell_{loc}$ is roughly $2a$ comparable with the surface distance
 $\ell_{surf}$ between nearest neighbor rods. 
Note generally that $\ell_{loc} \geq \ell_{mfp} \geq \ell_{surf}$ (usually 
 $\ell_{loc} \gg \ell_{mfp}$), where $\ell_{mfp}$ is the mean free path. 
Therefore, it would not be appropriate to conclude that Anderson localization
 can explain the observed localization length within the PGs. 
In fact, PGs have not been
 detected in two-dimensional disordered systems which show localization
 of light \cite{Localization}.
On the other hand, $\ell_{loc}$ above PGs is estimated to be $30a-40a$ comparable
 with the sample thickness.
The effect of localization is usually obvious
 when $\ell_{loc}$ is comparable with or less than the
 sample thickness $\ell$. 
Therefore, the thickness dependence of $T$ outside PG reflects the effect
 of Anderson localization.

There are numerous studies concerning the effect of disorder
 such as randomness of radii, positions or dielectric constants of rods
 on the PGs \cite{DisorderedPC} and waveguides made of PhCs \cite{Bayindir01}.
PGs are observed when disorder is not so strong, but they are obviously vestiges
 of PGs of the underlying lattices. 
In contrast, there is no underlying lattice for UDPS and no peak is observed
 in radial distribution functions $g(r)$ as shown in Fig. 2b. 
The discovery of UDPS has three important contributions. 
Firstly, it
 gives a conceptual breakthrough of the common belief that either periodicity
 or short-range order is indispensable for the existence of PGs. 
Secondly, it enables us to fabricate very easily the wavelength-sized
 optical waveguides. 
Lastly, it unveils a new and powerful role of well-controlled randomness
 which can drastically change the optical feature of photonic materials.
There has been no study, to the authors' knowledge, to
 recognize such an active role of randomness in photonic materials.
Future researches of air-hole type UDPS to make use of TE modes and better
 design policy to increase uniformity are needed for further development of UDPS. 

This research was supported by a Grant-in-Aid for Scientific Research from
 the Ministry of Education, Culture, Sports, Science, and Technology. 
One of the authors (H. M.) expresses his sincere thanks to R. Ohkawa for
 his continuous encouragement.
We are grateful to K. Ohtaka for valuable discussions.

\begin{figure} 
\caption{
Top view of distributions of rods under the condition $D_{min}=4.0a$ in (a), (b) and
 Transmittance $T$ in (c). 
Circles show rods of radius $a$ and $\varepsilon=12$ in the rectangular region 
 of width $W=84.6a$ and height $H=53.6a$ within the vacuum. 
Total number $N$ and area fraction $f$ of rods are $N=100$ and $f=0.069$ in (a) and
 $N=200$ and $f=0.138$ in (b). 
The electric field parallel to the rod axis (TM mode) of
 wavelength $\lambda$ is incident from the upper side of (a) or (b). 
$T$ is calculated as a function of $\Omega=2\pi a/\lambda$
 by averaging the energy flow at line L in (a) or (b). 
Values of $T$ for three cases of $N$=100, 150 and 200 are shown in (c). 
We also plot $T$ for UDPS with $N$=200 and radius fluctuation of
 $\Delta a/a=\pm 20\%$. 
Each value of $T$ is the average of 5 different configurations. 
Central frequency and width of PG for $N$=150 are $\Omega_c=0.398$
 and $\Delta \Omega=0.119$, respectively. 
For $N$=200, $\Omega_c=0.431$ and $\Delta \Omega=0.159$. 
Introduction of $\Delta a/a =\pm 20\%$ only reduces $\Delta \Omega$
 to $\Delta \Omega=0.128$.
}
\label{F1}
\end{figure}

\begin{figure}
\caption{
(a) Top view of the distribution of rods, (b) two examples of radial distribution
 function $g(r)$ and (c) $T$ for $D_{min}=2.1a$.
Here, $N=200$, $W=37.5a$ and $H=33.3a$ (area fraction $f=0.503$).
No radius fluctuation is introduced.
Horizontal and vertical axes of (b) are the distance $r$ between rod centers
 in units of $a$ and its frequency $g(r)$, respectively.
Two spectra indicated by $\alpha$ and $\beta$ in (c) are transmittance of the TM mode corresponding to $\alpha$ and $\beta$ in (b), 
 and the lowest one is the average transmittance over five configurations for the TE mode
 (${\bf E} \perp$ rod axis). 
Gaps of TM and TE modes are respectively given by $0.50 \leq \Omega \leq 0.58$,
 $0.87 \leq \Omega \leq 0.98$, $1.28 \leq \Omega \leq 1.39$ and
 $0.67 \leq \Omega \leq 0.69$.
}
\label{F2}
\end{figure}

\begin{figure} 
\caption{
Transmittance $T$ of thin UDPS as a function of sample thickness $\ell$. 
Rod density and incident light are the same as in Fig. 1b. 
No radius fluctuation is introduced. 
We show in the inset one example of thin UDPS with $\ell=20a$ defined by outer dotted lines. 
From this UDPS, two thinner UDPS are cut out at horizontal lines indicated by $\alpha$ or $\beta$, whose thickness is $\ell=6.7a$ or $13.3a$. 
Rods on the cutting line are included when their centers are above the line. 
$T$ of each UDPS is calculated by averaging the energy flow at line L of width $5.7a$
 and $2.9a$ below each cutting line. 
Each value of $T$ is the average over five different configurations. 
Central frequency and width of PG for $\ell=13.3a$ and $20a$ are
 respectively given as $\Omega_c=0.424$, $\Delta \Omega=0.113$ and 
 $\Omega_c=0.429$, $\Delta \Omega=0.146$. 
}
\label{F3}
\end{figure}

\begin{figure} 
\caption{
(Color).
Distributions of electric field intensity and energy flow in
 various waveguides made of UDPS. 
Field intensity increases from blue to red, and energy flow
 is indicated by the white arrows. 
Chosen frequencies in (a) and (c) are shown respectively by the black arrow
 in (b) and (d) which show transmittance $T$ obtained at line L. 
The shaded region is the PG. 
A waveguide (a) has $9.5a$ width and bends by $90^\circ$ with
 $N=229$, $W=72.0a$ and $H=81.3a$. 
A waveguide (c) is composed of two quarter circles with
 $N=238$, $W=106.7a$ and $H=53.3a$. 
Outer and inner radii of the circles are arbitrarily chosen
 as $31.4a$ and $21.9a$, respectively. 
Maximum field intensity in units of incident light is 15.5 in (a) and
 11.9 in (c). 
In all the cases, density of rods of UDPS is the same as that in Fig. 1b
 and radius fluctuation $\Delta a/a=\pm 20\%$ is introduced. 
Dielectric constant of rods and incident light are the same as those in Fig. 1.
}
\label{F4}
\end{figure}

\end{document}